\documentclass[twoside,twocolumn]{article}

\usepackage[sc]{mathpazo} 
\usepackage[T1]{fontenc} 
\usepackage{lmodern} 
\usepackage{microtype} 

\usepackage[english]{babel} 

\usepackage{graphicx}
\usepackage{color}
\usepackage[top=2cm,bottom=2cm,left=1.5cm,right=1.5cm,columnsep=20pt]{geometry} 
\newenvironment{compdetail}
    {\large\centering\scshape Computational Details%
    \par\medskip\normalfont\normalsize}%
    {}%
\newenvironment{acknowledgement}
    {\large\centering\scshape Acknowledgement%
    \par\medskip\normalfont\normalsize}%
    {}%
    {\large\centering\scshape Supporting Information%
    \par\medskip\normalfont\normalsize}%
    {}%
\newenvironment{author contributions}
    {\large\centering\scshape Author contributions%
    \par\medskip\normalfont\normalsize}%
    {}%
\newenvironment{competing interests}
    {\large\centering\scshape Competing interests%
    \par\medskip\normalfont\normalsize}%
    {}%
\newenvironment{figure legends}
    {\large\centering\scshape Figure legends%
    \par\medskip\normalfont\normalsize}%
    {}%
\usepackage{amsmath}

\usepackage[normal, small,labelfont=bf,up,textfont=it,up]{caption} 
\usepackage{booktabs} 

\usepackage{lettrine} 

\usepackage{enumitem} 
\setlist[itemize]{noitemsep} 

\usepackage{abstract} 

\usepackage{titlesec} 
\renewcommand\thesection{\Roman{section}} 
\renewcommand\thesubsection{\roman{subsection}} 
\titleformat{\section}[block]{\large\scshape\centering}{\thesection.}{1em}{} 
\titleformat{\subsection}[block]{\large}{\thesubsection.}{1em}{} 

\usepackage{fancyhdr} 
\usepackage{titling} 

\usepackage{hyperref} 

\pretitle{\begin{center}\Large\bfseries} 
\posttitle{\end{center}} 

\def\bea{\begin{eqnarray}}
\def\eea{\end{eqnarray}}
\def\ben{\begin{equation}}
\def\een{\end{equation}}
\def\benu{\begin{enumerate}}
\def\enu{\end{enumerate}}

\def\bei{\begin{itemize}}
\def\eei{\end{itemize}}
\def\beit{\begin{itemize}}
\def\eit{\end{itemize}}
\def\benu{\begin{enumerate}}
\def\enu{\end{enumerate}}

\def\sss{\scriptscriptstyle\rm}

\def\1var{(\bx_1...\bx\N)}

\def\bx{{x}}

\def\N{_{\sss N}}

\def\sph_int{ {\int d^3 r}}

\usepackage{multirow} 
\usepackage{slashbox} 

\title{Quantifying Density Errors in DFT}
\author{%
\textsc{Eunji Sim}\thanks{esim@yonsei.ac.kr} \textsc{and Suhwan Song} \\ 
\normalsize Department of Chemistry, Yonsei University, 50 Yonsei-ro Seodaemun-gu, Seoul 03722, Korea \\ 
\textsc{Kieron Burke} \\ 
\normalsize Departments of Chemistry and of Physics, University of California, Irvine, CA 92697, USA \\  
\and 
}
\date{\today} 

\renewcommand{\maketitlehookd}{%
\begin{abstract}
\noindent 
We argue that any general mathematical measure of density error, no matter
how reasonable, is too arbitrary to be of universal use.
However the energy functional itself provides a universal relevant measure 
of density errors.  For the self-consistent density of any Kohn-Sham
calculation with an approximate functional,
the theory of density-corrected density functional theory (DC-DFT) provides an accurate, practical estimate of
this ideal measure.
We show how to estimate the significance of the
density-driven error even when exact densities are unavailable.
In cases with large density errors, the amount of exchange-mixing
is often adjusted, but we show this is unnecessary.
Many chemically relevant examples are given.
\end{abstract}
}

\begin{document}

\maketitle






\sf
The Kohn-Sham (KS) approach to density functional theory (DFT) has become
very popular, being employed in more than 30,000 scientific papers each year\cite{PGB15}.
However, its claim to being first-principles is diluted by the fact that there
are literally hundreds of possible exchange-correlation
approximations available in most codes\cite{B12}.
Over many decades, many researchers have tried studying the self-consistent densities
of DFT calculations to gain insight into the quality of approximations\cite{C01,BG97}.
This interest recently intensified with the publication of Medvedev et al.\cite{MBSP17}, 
which appears
to show that the self-consistent densities of recent, empirically-parametrized
functionals are of poorer quality than those that are more systematically derived,
by careful comparison with accurate densities of atoms and ions.
But their conclusions depend on their choice of
how to measure the density error, which was designed to penalize incorrect oscillations
in approximate densities.  Other reasonable choices lead to
other conclusions.
Because the density is a function
and not a number, the variety of possible error-measures that people can
create is endless.
We give many alternative measures, some of which produce starkly different rankings.
Thus statements such as ``densities became closer to the exact ones" are entirely
dependent on the choice of metric and since every researcher can choose their own,
are of limited value at best.

However, the variational principle provides a natural and unambiguous measure of the accuracy
of a density for any systems.  Moreover, this metric is measured in terms of the resulting energy error. 
If errors in densities are so small that they have negligible impact on calculated
energies, they surely are not the most important indicator for
improving functional approximation.
The
well-established theory behind density-corrected DFT (DC-DFT) was
constructed precisely to measure this error, for self-consistent densities from
DFT calculations with approximate functionals.  Here we show how this can be done, or
at least usefully estimated, 
even in the absence of the exact density.  
By measuring density errors via the energy, we find no evidence that empirical
functionals yield greater density errors.

\begin{figure}[htb]
\includegraphics[scale=0.5]{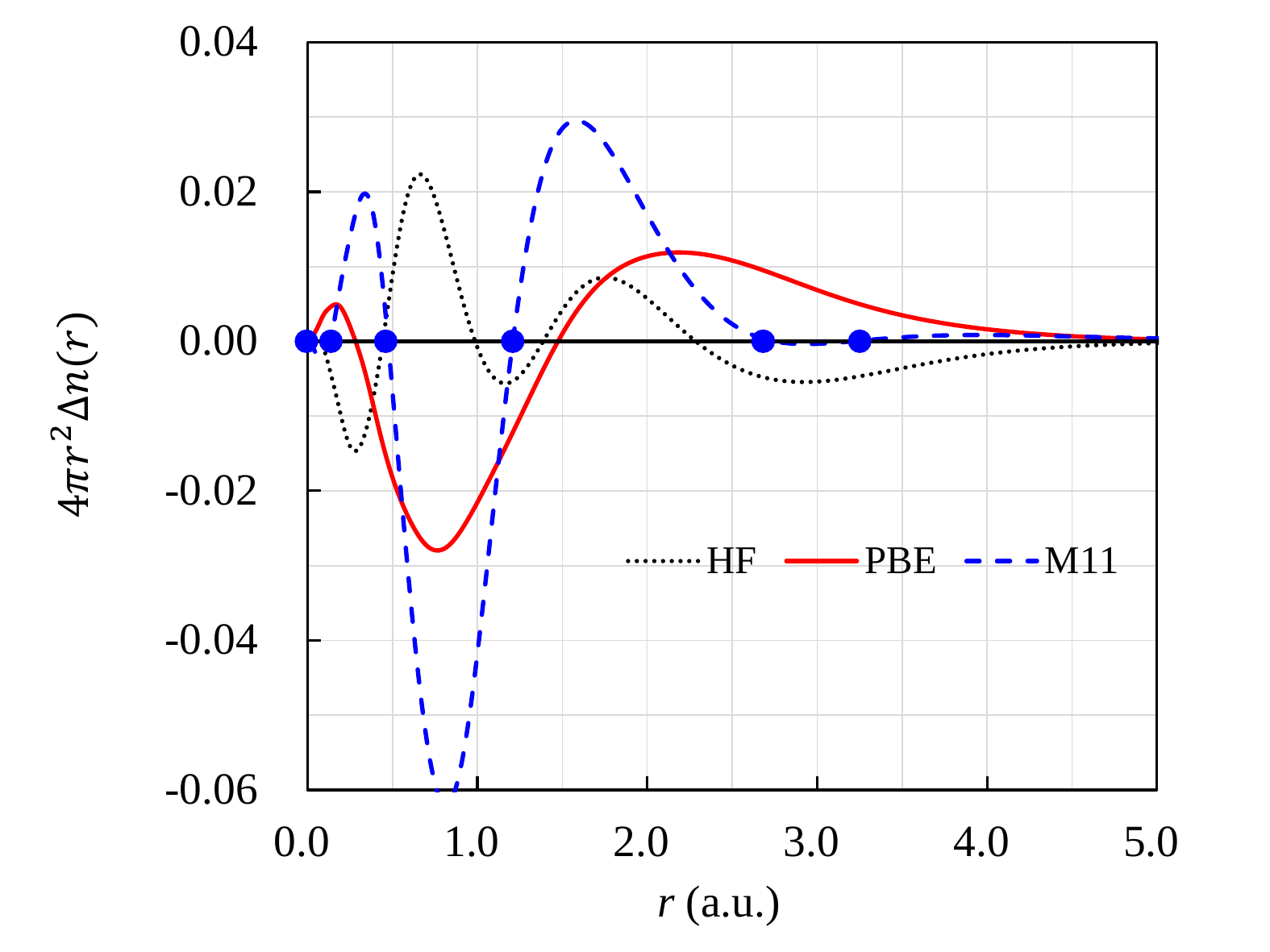}
\caption{He atomic radial density error for HF (gray-dotted), PBE (red-solid),
and M11 (blue-dashed), relative to highly accurate quantum Monte Carlo density.  Radii where the M11
density crosses the QMC density are marked with solid dots.}
\label{heradial}
\end{figure}
Begin with the general idea of measuring errors in densities.
While a single number characterizes the error in an approximate energy, the error
in an approximate density, $\tilde{n}(\bf r)$, can be quantified in infinitely many ways.
For example, an intuitively appealing measure is the simple L2 norm:
\begin{equation}
\Delta_{L2}[\tilde{n}] = \int d^3r\, (\tilde{n}(\bf{r})-\textit{n}(\textbf{r}))^2,
\label{L2}
\end{equation}
where $n(\bf r)$ is the exact density, and
$N=\int d^3r\ n(\bf r)$ is the number of electrons.
Ref.\cite{MBSP17}
looks at many different functionals (128), on a few
very simple systems (14 small closed-shell atoms and ions) and construct
three different measures (a discretized $\Delta$, and analogs for gradients and Laplacians)
that are combined in several ways to yield some measure of density error.
However, the relative ranking depends strongly on the measure.

To understand why such a definition of density error does not suffice, Fig.\ref{heradial}
shows the radial density error of three different approximations for the He atom.
The exact reference density is from a highly accurate quantum Monte Carlo (QMC) calculation\cite{UG94},
and all results are converged to the basis set limit.  Since we are only interested in general
features, 3 approximations suffice to make our point.  We show the error in the
Hartree-Fock (HF) density, that of
self-consistent
PBE, and that of M11, a relatively recent Minnesota functional.
By the measures of Ref.\cite{MBSP17}, M11 has substantially worse densities than PBE and even HF,
despite providing much better thermochemistry than the latter.

There are infinitely many possible choices of measure.  To give an idea of
the possible variety, note that the M11 density matches the exact density at 6 radial
points.  If we define our measure of error as 
\begin{equation}
\Delta_{RP}[\tilde{n}]=\sum_{i=1}^6 (\tilde{n}(r_i)-n(r_i))^2,
\label{point}
\end{equation}
where the sum is over the crossing points (marked by dots in Fig.\ref{heradial}), then M11 has
zero error, but all other approximations have finite errors (including PBE and HF in the figure),
i.e., it beats all 127 other functionals in the study, if we apply this strategy to the 14 systems
studied.  But we could equally as well have taken the PBE crossing points, making it the winner,
or
indeed {\em any} of the 128 candidates.
While this example may appear facetious, it
explicitly demonstrates that the rankings can be arbitrarily
reordered. We could even include a 129th, namely setting $E_{XC}=0$ (Hartree approximation), 
and have it
be the best.

One might claim that such measures are unreasonable, while using 
the ingredients
of semilocal functional approximations is more reasonable.  But the number of such combinations
is endless and other energy components, such as the KS kinetic energy or the Hartree energy,
are also used in a Kohn-Sham (KS) calculation.
In fact, the original approximate DFT, Thomas-Fermi theory, has been rigorously proven
to yield relatively exact energies for all systems\cite{LS77}, despite never producing the correct
density pointwise, in a certain limit of large particle number.  As that limit is approached,
the density develops an ever more rapid oscillation (shell structure) which such reasonable
measures penalize.
The same has been 
argued for the local density approximation (LDA) \cite{KS65}
within KS-DFT\cite{BCGP16}, and even the ionization
energies of atoms appear to become relatively exact in this limit\cite{CSPB10}. 
In all these cases,
the approximate functional derivatives (potentials) miss shell structure, with corresponding
oscillating errors in the densities (Fig.4 of Ref.\cite{CSPB10}). 

There have been many responses and follow-up analysis to Ref.\cite{MBSP17},
and we highlight only a few.
In a technical comment, Kepp shows that errors in approximate energies
and their corresponding approximate density errors (as defined
in Ref.\cite{K17c}) are largely uncorrelated and chemically irrelevant.  
Brorsen et al.\cite{BYPH17} come to a similar
conclusion, but studying specifically the density accumulation in a bonded region.
Wang et al.\cite{WWTH17} argue that, with a smaller, more standard
basis,
the recent Minnesota functionals rank more favorably.
This is heresy in terms of traditional quantum chemistry, as it argues
that an unconverged
calculation yields more useful results than a converged one.
But, since different measures evaluated
on different systems yield different rankings, why not this one?
In fact, the reply to Kepp's
comment beautifully summarizes the deep intellectual
argument for non-empirical functional approximation\cite{MBSP17K}, and the
recent successes of the SCAN meta-GGA confirm its practical significance\cite{SRP15},
but such arguments do not determine a metric.
While it is true that when
trying to evaluate the quality of an approximate density functional,
databases of different systems and properties are often used.  Some functionals
are designed only for molecular problems, while others also include materials.
All such evaluations require human choices of which systems to include and how to
weight the errors, i.e., prudent use of statistics.  But in the deterministic world of
electronic structure,
we should not resort to statistics unless we must.

In fact, there is a simple measure of density error that circumvents all these
difficulties.  To see this, consider regular quantum mechanics, where the variational
principle tells us, for a given system:
\begin{equation}
E = \min_\psi \langle \psi | \hat H | \psi \rangle,
\end{equation}
where $\hat H$ is the Hamiltonian, $\psi$ is any allowed normalized trial wavefunction,
and $E$ is the ground-state energy.  Thus we naturally say one wavefunction is better than
another, meaning it yields a lower energy.  The Hamiltonian itself provides an unambiguous
metric for the quality of a wavefunction for a specific system. The exact wavefunction is
always the winner, and no choices have been made.  Of course, a better wavefunction does not
mean better by every measure.  Moreover, the error is measured in precisely the units we typically
care most about, i.e., the energy.

\begin{figure}[htb]
\includegraphics[scale=0.5]{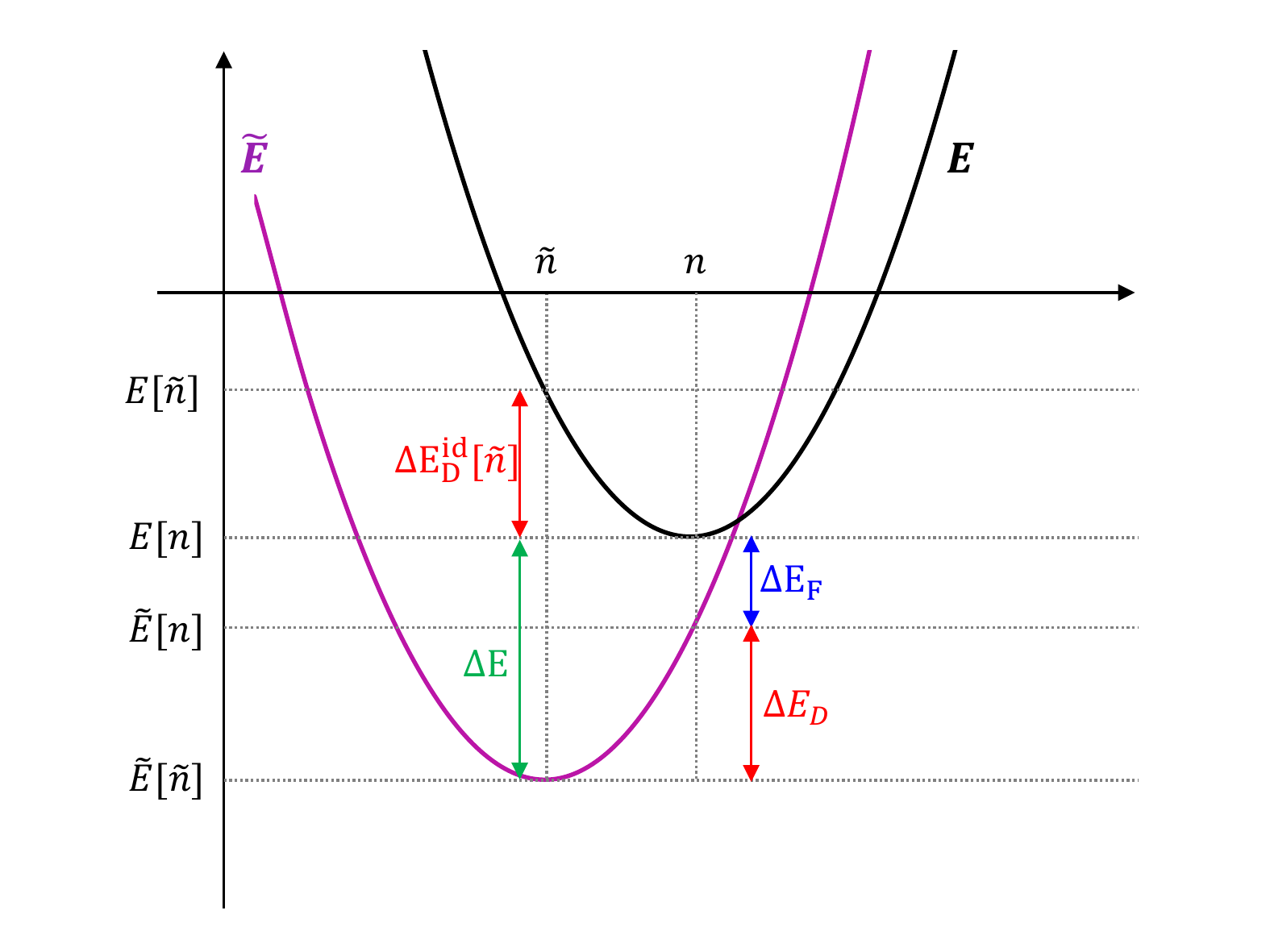}
\caption{
Cartoon of the exact (black) and approximate (purple, denoted with
tilde) energy functionals.  The minima are marked as $n$ and $\tilde n$, respectively.
Our $\Delta E_D$ mimics the ideal (but practically uncomputable) $\Delta E_D^{id}$,
the energy error of the exact functional
due to evaluation on $\tilde n$ instead of $n$. The calculation shown is abnormal.
(If $\tilde n$ is brought much closer to $n$, the red errors become much smaller
than the blue, and the system is normal.)}
\label{cartoon}
\end{figure}
The variational principle also applies to DFT.
For a given system,
determined by its one-body potential $v(\bf r)$, there is a well-defined, exact ground-state
energy $E_v$ and density, $n(\bf r)$. (For simplicity, we limit ourselves to
non-degenerate, spin-unpolarized
cases.)  
An ideal metric for the error in an approximate density $\tilde n$ is
\begin{equation}
\Delta E^{id}[\tilde{n}]= E[\tilde{n}]-E[n],
\label{Eid}
\end{equation}
where $E[n]$ is the exact total energy functional for a given system.  This is never negative
and, if it is much smaller than the corresponding energy error in 
the approximate functional that produced $\tilde{n}(\bf r)$, would
tell you that such density errors are unimportant.  In principle, this provides an unambiguous
relevant measure for all systems and approximate densities.
(Of course, for specific properties of practical interest, one might use
a different measure, such as the density at a nucleus
for NMR shifts.)

Unfortunately, evaluating $E[\tilde{n}]$ requires the
inversion of a many-body problem, which is more difficult than solving the original problem exactly,
and has only been achieved in a few model cases\cite{WSWB13}.  
However, in the special case of a self-consistent density from an approximate KS-DFT calculation,
we have an excellent proxy for $\Delta E^{id}[\tilde{n}]$, from the well-established theory of
DC-DFT.  
The true error in an XC approximation, $\tilde E_{XC}$, on any given density is
\begin{equation}
\Delta E_F[n] = \tilde E_{XC}[n]-E_{XC}[n] = \Delta E_{XC}[n],
\label{EF}
\end{equation}
called the functional error.
The density-driven error is defined as the difference between this and the total energy error:
\begin{equation}
\Delta E_D = \Delta E - \Delta E_{XC}[n] = \tilde E[\tilde{n}]-\tilde E[n],
\label{ED}
\end{equation}
where $\tilde E$ is the total energy functional when $\tilde E_{XC}$ is used.
As illustrated in Fig.\ref{cartoon}, 
if the approximate functional has the same curvature as the exact one and
the densities are sufficiently close, $\Delta E_D \approx -\Delta E^{id}[\tilde{n}]$. 
Practically, we can evaluate $\Delta E_D$ once we know the exact density and energy,
without needing $E[\tilde n]$.
Of all the 30,000+ DFT calculations published each year\cite{PGB15}, in any case where the
exact energy and density is known, the DFT errors can be decomposed in this way.

Examination of the ratio of the magnitudes of the functional- and density-driven
errors has shown that, in the vast majority of KS calculations, including all
those of Ref.\cite{MBSP17}, the error reported
in the energy is dominated by the functional error\cite{KSB13}.  
Such calculations are labelled {\em normal} in DC-DFT, and 
the density-driven error is a small
fraction of the total and so is 
irrelevant to the energy error.  
But in several important classes of DFT calculations  with standard functionals
(electron affinities, reaction barriers,
stretched heteronuclear bonds, and ions and radicals in solution, for example), density-driven
errors are significant.  These are labelled {\em abnormal}, and errors decrease
significantly if exact densities are used instead of 
self-consistent ones\cite{KSB14, KPSS15, WNJK17}.
In practice, it is rarely straightforward (or inexpensive) to generate highly accurate
densities (and hybrid functionals also require KS orbitals).  In practice, for molecules, 
HF densities usually suffice, because they do not suffer from self-interaction (or delocalization)
error as semilocal approximations do.  With a simple script\cite{TCCL} and at no
additional computational
cost, a HF-DFT calculation (i.e., DFT energy on HF density)
can be run and yields substantially improved energetics.

To illustrate our assertions, begin with simple one- and two-electron calculations, where
highly accurate densities are available, so that density-driven errors can be precisely calculated.
We choose an energy error standard of $\Delta_s = $ 2 kcal/mol, as we do not expect our approximations
to achieve that accuracy, and so regard errors smaller than $\Delta_s$ as not meaningful.
In Table ~\ref{tbl:1e2e}, we show energy errors for the humble H atom.  All approximations in the table have errors below
$\Delta_s$.  We also see that the true and density-driven errors are also below
threshold.  Ironically, the non-empirical PBE has almost exactly zero error in this case, but our 
decomposition show this is accidental, as the functional and density-driven errors cancel almost perfectly. 

\begin{table*}[htb]
\centering
\resizebox{15cm}{!}{
\begin{tabular}{|l|rrr|rrr|rrr|rrr|}
\hline
 \multirow{2}{*}{\backslashbox{System}{Error}}  & \multicolumn{3}{|c|}{PBE} & \multicolumn{3}{|c|}{PBE0} & \multicolumn{3}{|c|}{M06} & \multicolumn{3}{|c|}{MN15} \\
       & $\Delta E$    & $\Delta E_F$  & $\Delta E_D$  & $\Delta E$     & $\Delta E_F$  & $\Delta E_D$  & $\Delta E$    & $\Delta E_F$  & $\Delta E_D$  & $\Delta E$     & $\Delta E_F$  & $\Delta E_D$  \\ \hline
H      & 0.0   & 0.4    & -0.4   & -0.8   & -0.7   & -0.2   & -0.1  & 0.3    & -0.4   & 0.3    & 2.0    & -1.7   \\
He$^+$    & 3.9   & 4.3    & -0.4   & 2.0    & 2.3    & -0.2   & 0.2   & 0.6    & -0.3   & -3.8   & -2.0   & -1.8   \\
He     & 6.8   & 7.5    & -0.7   & 5.4    & 7.5    & -2.1   & -4.1  & -1.8   & -2.3   & -11.0  & -7.1   & -3.9   \\
H$^-$     & -6.5  & -1.1   & -5.4   & -1.4   & 4.3    & -5.7   & -1.9  & 4.0    & -5.9   & -2.1   & 2.8    & -5.0   \\
\hline
He (IP) & -2.9  & -3.2   & 0.3    & -3.3   & -5.2   & 1.9    & 4.3   & 2.3    & 2.0    & 7.2    & 5.1    & 2.1    \\
H (EA)  & 6.5   & 1.5    & 5.0    & 0.6    & -5.0   & 5.5    & 1.8   & -3.7   & 5.5    & 2.4    & -0.8   & 3.3   \\ \hline
\end{tabular}
}
\caption{Errors (kcal/mol) in DFT approximations for one- and two-electron systems. Two-electron systems are calculated with CASSCF density as for the reference and compared with reference QMC energies.}
\label{tbl:1e2e}
\end{table*}

Turn next to He$^+$.  The density-driven errors are unchanged (as the shape of the density barely
changes), but the greater dominance of exchange unbalances PBE, which is then improved upon by PBE0.
Next consider He, where now errors are noticeably bigger.  Most have reasonable density errors (except MN15),
and all errors are dominated by the XC error.  
But chemistry concerns energy differences, so at the bottom of Table ~\ref{tbl:1e2e}, 
we give the ionization potential (IP)
of He and the electron affinity (EA) of H.  We apply exactly the same formulas to energy differences as we have
used for total energies.
All He ionization potential calculations have small density errors, and the error in the
ionization potential is dominated by the XC error:  these are all normal calculations.  

\begin{figure}[htb]
\includegraphics[scale=0.5]{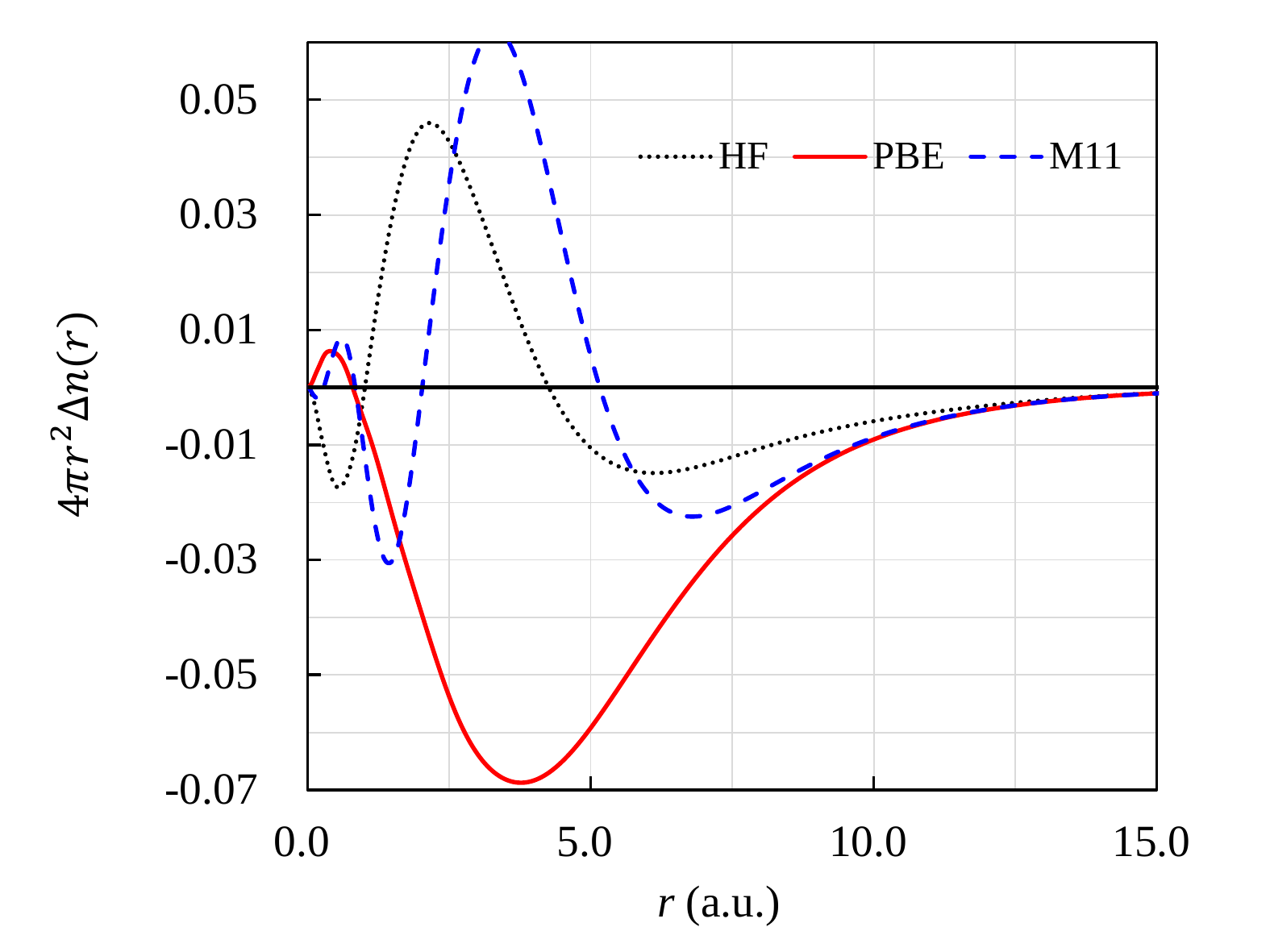}
\caption{H$^-$ radial density errors, for the same approximations
as in Fig.\ref{heradial}.}
\label{hradial}
\end{figure}

Next, we consider a paradigm of {\em abnormality}, namely the hydrogen anion. 
Its density errors are shown in Fig.\ref{hradial}.   Note the scale is larger than Fig.\ref{heradial}, 
and some errors do not oscillate.  With
many semilocal approximations, a self-consistent solution leads to about 0.35 electrons
escaping the system (the HOMO is then at 0, unlike in calculations where the anion
is bound artificially by the basis, and the HOMO is positive\cite{KSB13}).  
It is not the scale of the density errors that determines abnormality, but their
effect on the calculated energy.  Thus, H, He$^+$, and He are generically {\em normal}
systems with semilocal DFT, while atomic anions are generically {\em abnormal}\cite{LFB10,JD99}.

A crucial practical point is if abnormality in a system can be
detected without access to the exact density (if the exact $n(\bf r)$ were always needed,
there could be no practical benefit for larger systems).  
The many successes of HF-DFT\cite{WNJK17,SKSB18}
 show that, for almost all abnormal calculations, use of the HF density
instead of the exact density yields an excellent estimate of the density-driven error, yielding a very 
practical scheme for correcting those errors.  (The only exceptions are due to spin contamination of the
HF calculation\cite{KPSS15} or the ultra abnormal paradigm, H$^-$\cite{KSB13}.)  We thus
use this in our analysis of non-trivial abnormal systems.
Initial DC-DFT work showed
that, when the KS gap in a self-consistent semilocal DFT calculation
becomes unusually small, there is likely to be a substantial density-driven error\cite{KSB13}.
This is true for all cases listed above. 
However, this leaves open the question
of "how small is small enough?", i.e., it does not provide a clear quantitative
answer, and also does not use only energies of ground-state calculations.

The answer lies in measuring the sensitivity of the energy to different densities.
In recent work extending HF-DFT to the energy differences between spin states of
Fe(II) complexes, "rainbow" plots take
a range of different functionals of different classes, and show the
energy of every functional on every self-consistent density\cite{SKSB18}.  The spread in such a 
plot indicates the sensitivity to the density: If there is more variation with
density than there is with functional choice, then the system is abnormal.
Kepp took 4
popular approximations of different kinds and
estimated the degree of normality in a calculation\cite{K18} by
averaging over their evaluation on each other's self-consistent densities.
But this misses a key element in DC-DFT:  Separation of accuracy of densities and energies.
For example, the HF density yields much improved energetics with semilocal DFT energies, 
but the HF {\em energy} is so poor it is rarely used.  Thus Kepp's scheme does not
include pure HF densities.

To detect abnormality, we should measure the sensitivity of a functional to
the density.  So we evaluate on two extreme non-empirical densities: HF and LDA.
For any approximate functional,
define:
\begin{equation}
S(\tilde E_{XC})=\Bigl\lvert \tilde E[n^{\rm{LDA}}]- \tilde E[n^{\rm{HF}}] \Bigr\rvert,
\end{equation}
If $S > \Delta_s$, our 2 kcal/mol cutoff, this suggests
abnormality, which in turn implies a density-driven error that would be cured by using HF
densities (for molecules).  
Moreover, $S$ approximates the
density-driven error (see Eq.(\ref{ED})) if
the LDA density is close to the self-consistent density and the HF density
is close to the exact one. 
In principle, abnormality of a DFT calculation depends on the approximate functional
used, the system, and the energy difference being calculated.  However, abnormal
calculations with standard semilocal approximations are often much improved by use of HF densities.
We calculate $\bar S_m$, an average over $m$
approximate XC functionals.  Here, we chose $S_1$ as using just PBE, but $\bar S_3$ averaging
over LDA, PBE, and BLYP (i.e., generalized gradient approximation (GGA) level), and $\bar S_4$ as averaging over PBE, PBE0,
BLYP, and B3LYP, i.e., semilocal and hybrid level.
Table~\ref{tbl:many} shows results for
many different chemical processes.  While there is some spread in the different 
measures, there is great consistency:  Whenever the density sensitivity is above
2 kcal/mol, the system is abnormal at the semilocal/hybrid level.  
Atomization energies, standard reaction energies, and ionization energies are
all normal.  In such cases, use of the HF density barely changes the result, and
may not improve it.  In normal calculations, there is no reason to think the HF
density is more accurate than the self-consistent DFT density (and many reasons
to think it is not\cite{BG97}).  All entries for the systems of Ref.\cite{MBSP17}
indicate their normality, and their density-driven errors are very small.

\begin{table*}[htb]
\centering
\resizebox{15cm}{!}{%
\begin{tabular}{|l|crrr|cc|rr|rr|rr|rr|}
\hline
\multirow{2}{*}{\backslashbox[30mm]{System}{}}  & \multirow{2}{*}{$^a$} & \multirow{2}{*}{$S_1$}     & \multirow{2}{*}{$\bar S_3$}     & \multirow{2}{*}{$\bar S_4$}     & \multirow{2}{*}{$\mathbb{N}^b$}     & \multirow{2}{*}{$\bar S_{K}$} &\multicolumn{8}{c|}{DFT Errors$^c$} \\ 
  &  &  &     &     &      &   &PBE&[HF]&{B3LYP}&[HF]& M06 & [HF] & MN15 & [HF] \\ \hline
& \multicolumn{14}{c|}{Atomization energy} \\														
HCl	& {N}	& 0.2	& 0.3	& 0.1	& 0.1	& 0.4	& 0.5	& 0.4	& 1.8	& 1.7	& 0.6	& -0.1	& -1.0	& -1.7 \\
LiH	& {N}	& 0.6	& 0.5	& 0.4	& 0.1	& 0.4	& 4.8	& 5.3	& -0.2	& 0.1	& 0.1	& 0.5	& -5.6	& -6.7 \\
NaCl	& {N}	& 0.7	& 0.6	& 0.4	& 0.2	& 0.9	& 3.2	& 3.7	& 4.9	& 4.8	& -6.8	& -4.4	& -2.7	& -1.5 \\
CH$_2$($^3B_1$)	& {N}	& 0.8	& 1.4	& 1.1	& 0.3	& 1.2	& -2.1	& -1.2	& 0.1	& 1.6	& 0.1	& 1.5	& 4.3	& 3.3 \\
CH	& {N}	& 1.0	& 1.2	& 0.9	& 0.1	& 1.0	& -1.2	& -0.2	& -1.8	& -0.9	& -0.6	& 0.2	& 1.8	& 1.9 \\
CH$_2$	& {N}	& 1.1	& 1.2	& 0.9	& 0.1	& 1.0	& 3.3	& 4.4	& 1.2	& 2.1	& -0.2	& 0.8	& 3.9	& 1.4 \\
NH	& {N}	& 1.2	& 1.8	& 1.2	& 0.2	& 2.0	& -5.9	& -4.5	& -5.5	& -4.1	& 0.1	& 2.8	& 1.0	& 2.0 \\
CH$_4$	& {N}	& 1.4	& 1.6	& 1.1	& 0.1	& 1.6	& 0.9	& 2.2	& -0.2	& 1.0	& 2.0	& 4.0	& 3.3	& -0.7 \\
\hline														
& \multicolumn{14}{c|}{Reaction energy} \\														
HCl+CH$_3$$\rightarrow$CH$_4$+Cl	& {N}	& 0.5	& 0.6	& 0.2	& 0.2	& 0.7	& 1.8	& 2.3	& -0.6	& -0.3	& -0.1	& 0.8	& 1.7	& 1.2 \\
\hline														
& \multicolumn{14}{c|}{Ionization energy} \\														
C$_2$H$_2$	& {N}	& 0	.0& 0.1	& 0.1	& 0.0	& 0.1	& 2.1	& 2.0	& 2.0	& 2.0	& 5.7	& 6.0	& 2.6	& 2.2 \\
He	& {N}	& 0.4	& 0.5	& 0.3	& 0.3	& 0.3	& 2.6	& 3.1	& -8.4	& -7.9	& 0.6	& -0.1	& -1.0	& -1.7 \\
\hline
& \multicolumn{14}{c|}{Double Ionization energy} \\														
Ne$^{6+}$/Ne$^{8+}$	& {N}	& 0.4	& 0.8	& 0.3	& 0.2	& 0.6	& -21	& -22	& -14	& -15	& -66	& -67	& -37	& -44 \\
B$^{+}$/B$^{3+}$	& {N}	& 0.5	& 0.7	& 0.4	& 0.2	& 0.6	& -3.9	& -4.6	& 3.5	& 3.0	& -21	& -21	& -16	& -18 \\
N$^{3+}$/N$^{5+}$	& {N}	& 0.5	& 0.7	& 0.4	& 0.2	& 0.6	& -11	& -11	& -2.0	& -2.5	& -38	& -38	& -24	& -28 \\
O$^{4+}$/O$^{6+}$	& {N}	& 0.4	& 0.7	& 0.3	& 0.2	& 0.6	& -14	& -15	& -5.7	& -6.2	& -47	& -47	& -28	& -33 \\
\hline														
& \multicolumn{14}{c|}{Electron affinity} \\														
H	& {A}	& 8.5	& 10& 5.8	& 0.7	& 6.7	& 6.7	& -1.6	& 4.6	& 3.3	& 2.0	& -2.6	& 2.6	& 1.3 \\
\hline														
& \multicolumn{14}{c|}{Reaction barrier height} \\														
H+HF$\rightarrow$HF+H	& {A}	& 3.4	& 3.9	& 2.9	& 0.2	& 3.7	& -15	& -11	& -11	& -7.7	& -1.5	& 2.9	& -5.0	& -2.6 \\
CH$_4$+Cl$\rightarrow$HCl+CH$_3$	& {A}	& 3.7	& 4.1	& 2.9	& 0.2	& 3.2	& -7.8	& -4.1	& -1.1	& 1.5	& -0.6	& 2.0	& 0.3	& 5.6 \\
HCl+CH$_3$$\rightarrow$CH$_4$+Cl	& {A}	& 4.2	& 4.7	& 3.1	& 0.2	& 3.9	& -6.0	& -1.8	& -1.7	& 1.2	& -0.7	& 2.7	& 2.0	& 6.7 \\
H+N$_2$O$\rightarrow$OH+N$_2$	& {A}	& 11	& 13	& 9.9	& 0.1	& 12	& -8.0	& 3.5	& -6.6	& 2.8	& 0.3	& 11	& -2.3	& 7.0 \\
OH+N$_2$$\rightarrow$H+N$_2$O	& {A}	& 20	& 21	& 16	& 0.7	& 18	& -29	& -9.2	& -8.8	& 5.9	& -10	& 5.3	& -6.2	& 7.4 \\
\hline														
& \multicolumn{14}{c|}{Dissociation of stretched molecule} \\														
H$_2^+$ (at 5.0 $\AA$)	& {A}	& 2.8	& 2.3	& 2.5	& 0.1	& 2.1	& -50	& -47	& -41	& -39	& -41	& -39	& -36	& -35 \\
NaCl (at 6.0 $\AA$)	& {A}	& 14	& 13	& 10	& 2.0	& 8.2	& -10	& 3.5	& -3.7	& 4.2	& -13	& -9.6	& -4.1	& -5.2 \\
NaCl (at 10.0 $\AA$)	& {A}	& 28	& 28	& 23	& 2.0	& 20	& -24	& 3.9	& -15	& 4.5	& -21	& -9.4	& -9.4	& -5.3 \\
\hline														
& \multicolumn{14}{c|}{Radical reaction energy} \\														
NH$_2$+H$\rightarrow$NH$_3$	& {N}	& 0.5	& 0.7	& 0.3	& 0.2	& 0.8	& -0.8	& -0.1	& -0.6	& -0.1	& 0.4	& 1.6	& 2.5	& 4.7 \\
NHCH$_3$+H$\rightarrow$NH$_2$CH$_3$	& {A}	& 2.1	& 2.8	& 1.4	& 0.4	& 2.0	& -3.8	& -1.3	& -2.7	& -1.1	& -1.1	& 0.8	& 0.5	& 3.7 \\
H+N$_2$O$\rightarrow$OH+N$_2$	& {A}	& 8.1	& 8.3	& 6.3	& 0.5	& 6.1	& -21	& -13	& -2.2	& 3.2	& -10	& -6.1	& -3.8	& 0.4 \\
\hline														
& \multicolumn{14}{c|}{High- and Low-spin energy difference} \\														
$[$Fe(NCH)$_6]^{2+}$	& {A}	& 44	& 43	& 28	& 3.4	& 32	& 48	& 3.7	& 24	& 2.2	& 14	& -11	& 24	& -2.0 \\
\hline 
\end{tabular}
}
\caption{Classification of $^a$normal (N) versus abnormal (A) calculations ($S$ \textgreater 2 kcal/mol).
In all cases, classification agrees with previous observations of abnormality.
$^b$Kepp's $\mathbb{N}$
function\cite{K18} should be above 1 kcal/mol for abnormal systems, but often is not.
$^c$CCSD(T) value was used for the reference value to evaluate the DFT errors.
Note that the basis set information is classified in the computational details section.
}
\label{tbl:many}
\end{table*}


But transition-state barriers\cite{S93,JS08,VPB12}, spin energy differences
of Fe(II) complexes\cite{SKSB18},  dissociation of 
stretched molecules\cite{KPSS15}, and some (but not all)
radical reaction energies\cite{KSB14} are abnormal.
In almost every abnormal case, 
use of a HF density
in place of the self-consistent density improves the energetics, often substantially.
This is an illustration of the greater accuracy of the HF density in such cases
by our measure.
Because of  averaging only self-consistent densities, Kepp's measure, 
precisely as given in Ref.\cite{K18}, is too insensitive
to indicate many abnormal cases (although altering the cutoff might improve its performance).
However, $S_K$, using Kepp's functionals but on LDA versus HF densities, does work.

\begin{figure}[htb]
\includegraphics[scale=0.5]{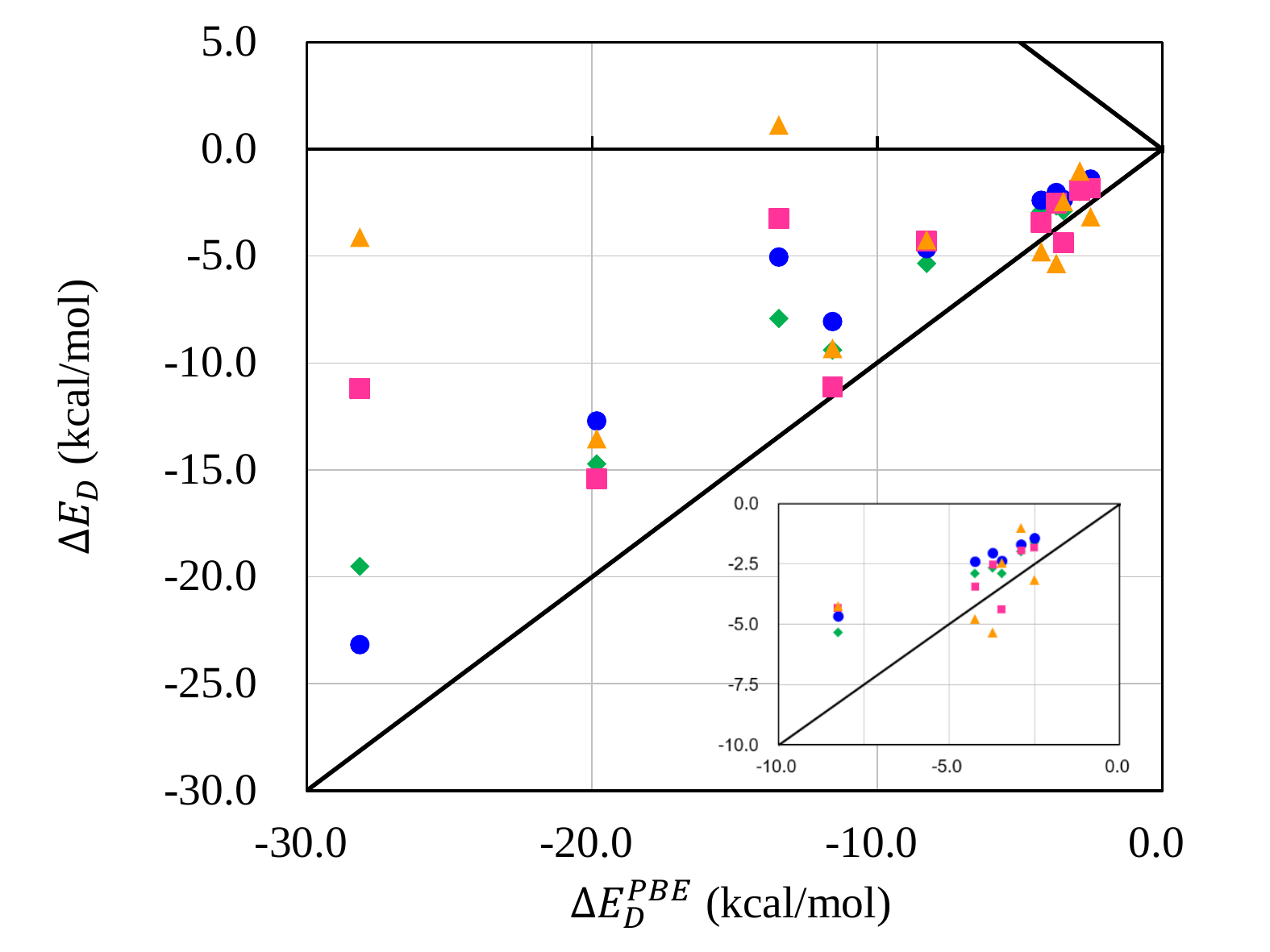}
\caption{Comparison of density-driven errors (calculated with HF densities
as the reference) for abnormal systems
of different functionals
against PBE in Table~\ref{tbl:many}.
B3LYP is labeled as green diamond, PBE0 for the blue dot, M06 for the pink square box, and MN15 for the orange triangle.
All points between diagonals (solid
black lines) have density-driven errors smaller than those of PBE.  
The electron affinity of H$^-$ has been excluded.
}
\label{ddevsdde}
\end{figure}

To illustrate the use of our measure in the present context, we consider the broad range
of processes covered by Table~\ref{tbl:many}.  Those labelled normal have density errors that 
are irrelevant by our measure: they do not matter to the accuracy of predicted energies.
But our abnormal systems are, by definition,
those in which the density error {\em is} relevant.  In Fig.\ref{ddevsdde}, we plot the density-driven
errors of several functionals against the density-driven error of PBE for all abnormal
processes in Table~\ref{tbl:many}.  Even in these extreme cases, we see no clear evidence of particularly
large density errors for specific classes of functionals.  Moreover, the 
mean absolute error (MAE) of these
approximations over the abnormal systems is 9.1 (kcal/mol) for PBE, 6.1 for PBE0, 5.4 for B3LYP, 5.3 for
M06, and 4.7 for MN15, again with no obvious pattern. (The inset suggests that, for abnormal
calculations with density-driven errors less than 10 kcal/mol, the empirical functionals may be
trading density-driven errors against functional errors, as one might expect from fitting data.)

Our last topic concerns the amount of exchange mixing in a global hybrid functional:
\begin{equation}
E\rm{hyb}[n]= a (E_X^{\rm{HF}}[n]-E_X^{\rm{GGA}}[n])+E_{XC}^{\rm{GGA}}[n]
\label{hyb}
\end{equation}
Becke introduced the idea and the
B3LYP functional uses $a=0.20$\cite{B93}.  Using arguments
from perturbation theory, Perdew et al. argued\cite{PEB96,BEP97,EPB97} $a=1/4$, the 
value used in PBE0\cite{ES99}.
A crucial aspect of such hybrids is that the amount of exchange mixing is fixed
in the functional, and not system-dependent, to retain size-consistency.
Moreover, large values of $a$ typically yield
highly inaccurate ground-state energetics (almost as bad as HF).
On the other hand, for many properties related to orbital eigenvalues, orbital-dependent
functionals are more accurate. The hybrid HSE06 often yields better gaps for semiconductors
within generalized KS calculations, and this can be crucial when studying localized
impurity levels\cite{HSE03}.  Recent papers often adjust 
$a$ to improve the positions of orbital energies.
However, such adjustments will destroy the quality (and generality) of the ground-state
energy functional but, worse still, create non-size-consistent schemes with ambiguities in
even the definition of the ground-state energy.  Thus, there is a parameter dilemma:
Increasing $a$ often improves response properties, but destroys the accuracy of
ground-state energetics and even
the rigor of the calculation.

\begin{figure}[htb]
\includegraphics[scale=0.5]{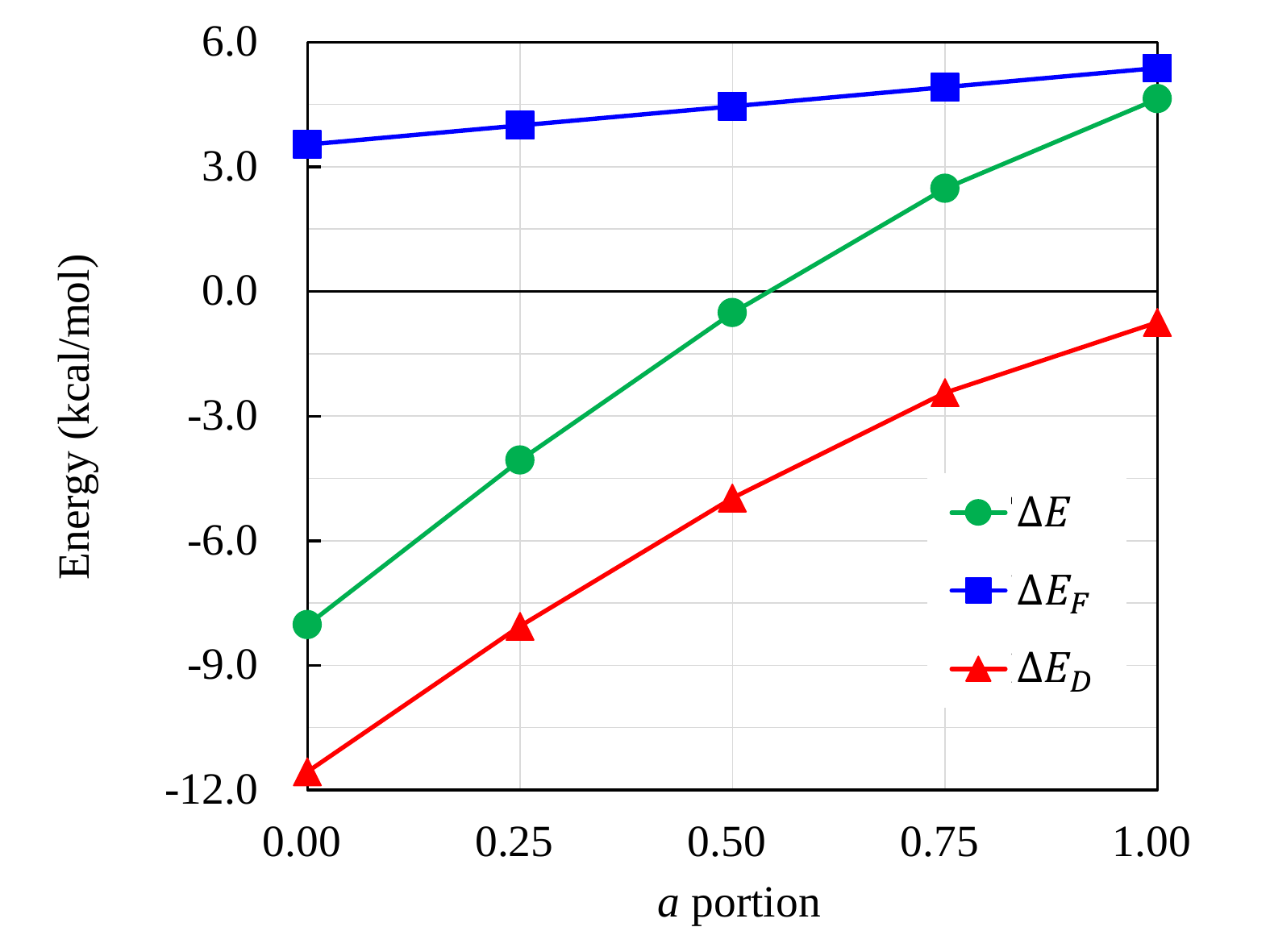}
\caption{Density-driven and functional errors for $a$-PBE calculation of abnormal
reaction barrier $H+N_2O \rightarrow OH+N_2$, as a function of amount of exact exchange
mixing (calculated with HF densities as the reference).
}
\label{adep}
\end{figure}

Here we show how HF-DFT avoids this dilemma, because the amount of mixing mostly affects
the quality of the density, not the energy.  In Fig.\ref{adep}, we see that, for
a typical abnormal calculation (a reaction barrier), the total energy error is very
sensitive to $a$ because the density-driven error is.  But the functional error barely
changes with $a$. 

We close with some discussion of context.   
The fact that the exact functional recovers the exact energy and density
tells us nothing about the accuracy of approximate functionals, for either the
energy or the density.  The concept of accuracy of approximations requires a quantitative measure
of error.  For any given system, an approximate functional makes an error in the
ground-state energy, but no single number can characterize all errors in the density.
The choices of Ref.\cite{MBSP17}, while emminently reasonable, are arbitrary, and any
subsequent ranking of approximations can be reordered with different choices.
In the most extreme case, any approximation can be made the best.

The energy functional itself provides a natural universal measure of density error, which
is easily estimated if a density is the result of an approximate KS calculation.
Moreover, it has led to the concept of abnormal DFT calculations, i.e., those in which
the density error contributes significantly to the calculated energy, and
where results can often be improved by using a `better' density.
In many (but not all) abnormal molecular calculations, the HF density suffices,
yielding the extremely trivial trick of performing HF-DFT calculations.
We note that HF-DFT is not a panacea, as it gives up the use of self-consistency.
Apart from the practical difficulties of dealing with the failure of the Hellmann-Feynman\cite{VPB12}
theorem, and while HF-DFT often improves energetics by a factor of 2 or 3 for
well-founded reasons, it does not apply to all delocalization errors, and
can never replace the need to get both energy and density accurately from a 
single functional.  The work of Bartlett and co-workers, called ab-initio DFT,
yields acccurate XC potentials and hence densities, while retaining self-consistency\cite{BLS05}.
Similarly, the work of Yang and co-workers to develop locally-scaled corrections
that restore the linearity of the energy with particle number produces
approximate functionals that solve these problems self-consistently.\cite{LZSY17}\\

\begin{compdetail}
All HF, DFT (SVWN\cite{D30, VWN80}, PBE\cite{PBE96}, B3LYP\cite{SDC94}, PBE0\cite{BEP97}, M06\cite{ZT08}, M11\cite{PT11}, MN15\cite{HHLT16})  HF-DFT, and CCSD(T) results are performed with Gaussian16 package \cite{G16}. 
Dunning's augmented correlation-consistent quintuple zeta basis set (aug-cc-pV5Z) is used for the calculations in Table~\ref{tbl:1e2e}\cite{D89, WD94}. For Table~\ref{tbl:many}, aug-cc-pVTZ basis set\cite{KDH92, WD93} was used for almost all calculation while Ahlrichs' newer redefinition quadruple zeta (def2-QZVP) basis set\cite{WFA03} is used for the dissociation of stretched molecules for the comparison with the previous work\cite{KPSS15}. To perform every calculation at their given orientation, molecular symmetry within the calculation was not considered.
Molecular geometry data are from the G2 database set\cite{RTSN02} for the atomization energy and ionization energy while molecular reaction geometries are from Zhao et al.\cite{ZLT05, ZGT05}. The high and low spin [Fe(NCH)$_6]^{2+}$ geometries are from Ref.\cite{SKSB18}. 
For the radical reaction energies, B3LYP/aug-cc-pVTZ is used for the geometry optimization.
Since the hydrogen anion of Tables~\ref{tbl:1e2e} and ~\ref{tbl:many} has the positive HOMO problem, SVWN, PBE, B3LYP, PBE0, and M06 calculations are performed with TURBOMOLE package\cite{T15} for calculating the fractional occupation within unrestricted KS scheme.
For the energy convergence criteria, SCF=tight option for the Gaussian16 while scfconv=8 and denconv=1.0d-8 are used for TURBOMOLE.
To calculate errors in Table~\ref{tbl:1e2e}, DFT exchange-correlation energies for two-electron systems are calculated on the CASSCF density from 60 active spaces and 2 active electrons\cite{pyscf} and compared with QMC energies\cite{UG94}. 
\end{compdetail}

\vspace{5mm}
\begin{acknowledgement}
This work at Yonsei University was supported by the grant from the Korean Research Foundation (2017R1A2B2003552).
KB acknowledges NSF CHEM 1464795.
\end{acknowledgement}

\vspace{5mm}
\begin{author contributions}
E.S. and K.B. designed research. S.S. performed
research. E.S., S.S., and K.B. analyzed data.
The manuscript was written by K.B., E.S., and S.S.
\end{author contributions}

\vspace{5mm}
\begin{competing interests}
The authors declare no competing interests.
\end{competing interests}

\clearpage


\begin{thebibliography}{10}

\bibitem{PGB15}
Aurora Pribram-Jones, David~A. Gross, and Kieron Burke.
\newblock Dft: A theory full of holes?
\newblock {\em Annu. Rev. Phys. Chem.}, 66(1):283--304, 2015.

\bibitem{B12}
K.~Burke.
\newblock Perspective on density functional theory.
\newblock {\em J. Chem. Phys.}, 136:150901, 2012.

\bibitem{C01}
Dieter Cremer.
\newblock Density functional theory: coverage of dynamic and non-dynamic
  electron correlation effects.
\newblock {\em Mol. Phys.}, 99(23):1899--1940, 2001.

\bibitem{BG97}
Evert~Jan Baerends and Oleg~V Gritsenko.
\newblock A quantum chemical view of density functional theory.
\newblock {\em J. Phys. Chem. A}, 101(30):5383--5403, 1997.

\bibitem{MBSP17}
Michael~G Medvedev, Ivan~S Bushmarinov, Jianwei Sun, John~P Perdew, and
  Konstantin~A Lyssenko.
\newblock Density functional theory is straying from the path toward the exact
  functional.
\newblock {\em Science}, 355(6320):49--52, 2017.

\bibitem{UG94}
Cyrus~J Umrigar and Xavier Gonze.
\newblock Accurate exchange-correlation potentials and total-energy components
  for the helium isoelectronic series.
\newblock {\em Phys. Rev. A}, 50(5):3827, 1994.

\bibitem{LS77}
Elliott~H Lieb and Barry Simon.
\newblock The {T}homas-{F}ermi theory of atoms, molecules and solids.
\newblock {\em Adv. Math.}, 23(1):22 -- 116, 1977.

\bibitem{KS65}
Walter Kohn and Lu~Jeu Sham.
\newblock Self-consistent equations including exchange and correlation effects.
\newblock {\em Phys. Rev.}, 140(4A):A1133, 1965.

\bibitem{BCGP16}
Kieron Burke, Antonio Cancio, Tim Gould, and Stefano Pittalis.
\newblock Locality of correlation in density functional theory.
\newblock {\em J. Chem. Phys.}, 145(5):054112, 2016.

\bibitem{CSPB10}
Lucian~A. Constantin, John~C. Snyder, John~P. Perdew, and Kieron Burke.
\newblock Communication: Ionization potentials in the limit of large atomic
  number.
\newblock {\em J. Chem. Phys.}, 133(24):241103, 2010.

\bibitem{K17c}
Kasper~P Kepp.
\newblock Comment on `'density functional theory is straying from the path
  toward the exact functional''.
\newblock {\em Science}, 356(6337):496--496, 2017.

\bibitem{BYPH17}
Kurt~R Brorsen, Yang Yang, Michael~V Pak, and Sharon Hammes-Schiffer.
\newblock Is the accuracy of density functional theory for atomization energies
  and densities in bonding regions correlated?
\newblock {\em J. Phys. Chem. Lett.}, 8(9):2076--2081, 2017.

\bibitem{WWTH17}
Ying Wang, Xianwei Wang, Donald~G Truhlar, and Xiao He.
\newblock How well can the m06 suite of functionals describe the electron
  densities of ne, ne$^{6+}$, and ne$^{8+}$?
\newblock {\em J. Chem. Theory Comput.}, 13(12):6068--6077, 2017.

\bibitem{MBSP17K}
Michael~G Medvedev, Ivan~S Bushmarinov, Jianwei Sun, John~P Perdew, and
  Konstantin~A Lyssenko.
\newblock Response to comment on ?쐂ensity functional theory is straying from
  the path toward the exact functional??
\newblock {\em Science}, 356(6337):496--496, 2017.

\bibitem{SRP15}
Jianwei Sun, Adrienn Ruzsinszky, and John~P Perdew.
\newblock Strongly constrained and appropriately normed semilocal density
  functional.
\newblock {\em Phys. Rev. Lett.}, 115(3):036402, 2015.

\bibitem{WSWB13}
Lucas~O. Wagner, E.~M. Stoudenmire, Kieron Burke, and Steven~R. White.
\newblock Guaranteed convergence of the kohn-sham equations.
\newblock {\em Phys. Rev. Lett.}, 111:093003, Aug 2013.

\bibitem{KSB13}
Min-Cheol Kim, Eunji Sim, and Kieron Burke.
\newblock Understanding and reducing errors in density functional calculations.
\newblock {\em Phys. Rev. Lett.}, 111(7):073003, 2013.

\bibitem{KSB14}
Min-Cheol Kim, Eunji Sim, and Kieron Burke.
\newblock Ions in solution: Density corrected density functional theory
  (dc-dft).
\newblock {\em J. Chem. Phys.}, 140(18):18A528, 2014.

\bibitem{KPSS15}
Min-Cheol Kim, Hansol Park, Suyeon Son, Eunji Sim, and Kieron Burke.
\newblock Improved dft potential energy surfaces via improved densities.
\newblock {\em J. Phys. Chem. Lett.}, 6(19):3802--3807, 2015.

\bibitem{WNJK17}
Adam Wasserman, Jonathan Nafziger, Kaili Jiang, Min-Cheol Kim, Eunji Sim, and
  Kieron Burke.
\newblock The importance of being self-consistent.
\newblock {\em Annu. Rev. Phys. Chem.}, 68(1):555--581, 2017.

\bibitem{TCCL}
\url{http://tccl.yonsei.ac.kr}.

\bibitem{LFB10}
Donghyung Lee, Filipp Furche, and Kieron Burke.
\newblock Accuracy of electron affinities of atoms in approximate density
  functional theory.
\newblock {\em J. Phys. Chem. Lett.}, 1(14):2124--2129, 2010.

\bibitem{JD99}
Andrzej~A Jarecki and Ernest~R Davidson.
\newblock Density functional theory calculations for f-.
\newblock {\em Chem. Phys. Lett.}, 300(1-2):44--52, 1999.

\bibitem{SKSB18}
Suhwan Song, Min-Cheol Kim, Eunji Sim, Anouar Benali, Olle Heinonen, and Kieron
  Burke.
\newblock Benchmarks and reliable dft results for spin gaps of small ligand fe
  (ii) complexes.
\newblock {\em J. Chem. Theory Comput.}, 14(5):2304--2311, 2018.

\bibitem{K18}
Kasper~P Kepp.
\newblock Energy vs. density on paths toward more exact density functionals.
\newblock {\em Phys. Chem. Chem. Phys.}, 20(11):7538--7548, 2018.

\bibitem{S93}
Jorge~M Seminario.
\newblock Energetics using dft: comparions to precise ab initio and experiment.
\newblock {\em Chem. Phys. Lett.}, 206(5-6):547--554, 1993.

\bibitem{JS08}
Benjamin~G Janesko and Gustavo~E Scuseria.
\newblock Hartree--fock orbitals significantly improve the reaction barrier
  heights predicted by semilocal density functionals.
\newblock {\em J. Chem. Phys.}, 128(24):244112, 2008.

\bibitem{VPB12}
Prakash Verma, Ajith Perera, and Rodney~J. Bartlett.
\newblock Increasing the applicability of dft i: Non-variational correlation
  corrections from hartree--fock dft for predicting transition states.
\newblock {\em Chem. Phys. Lett.}, 524:10 -- 15, 2012.

\bibitem{B93}
Axel~D Becke.
\newblock Density-functional thermochemistry. iii. the role of exact exchange.
\newblock {\em J. Chem. Phys.}, 98(7):5648--5652, 1993.

\bibitem{PEB96}
John~P. Perdew, Matthias Ernzerhof, and Kieron Burke.
\newblock Rationale for mixing exact exchange with density functional
  approximations.
\newblock {\em J. Chem. Phys.}, 105(22):9982--9985, 1996.

\bibitem{BEP97}
Kieron Burke, Matthias Ernzerhof, and John~P Perdew.
\newblock The adiabatic connection method: a non-empirical hybrid.
\newblock {\em Chem. Phys. Lett.}, 265(1-2):115--120, 1997.

\bibitem{EPB97}
Matthias Ernzerhof, John~P Perdew, and Kieron Burke.
\newblock Coupling-constant dependence of atomization energies.
\newblock {\em Int. J. Quantum Chem.}, 64(3):285--295, 1997.

\bibitem{ES99}
Matthias Ernzerhof and Gustavo~E Scuseria.
\newblock Assessment of the perdew--burke--ernzerhof exchange-correlation
  functional.
\newblock {\em J. Chem. Phys.}, 110(11):5029--5036, 1999.

\bibitem{HSE03}
Jochen Heyd, Gustavo~E Scuseria, and Matthias Ernzerhof.
\newblock Hybrid functionals based on a screened coulomb potential.
\newblock {\em J. Chem. Phys.}, 118(18):8207--8215, 2003.

\bibitem{BLS05}
Rodney~J Bartlett, Victor~F Lotrich, and Igor~V Schweigert.
\newblock Ab initio density functional theory: The best of both worlds?
\newblock {\em J. Chem. Phys.}, 123(6):062205, 2005.

\bibitem{LZSY17}
Chen Li, Xiao Zheng, Neil~Qiang Su, and Weitao Yang.
\newblock Localized orbital scaling correction for systematic elimination of
  delocalization error in density functional approximations.
\newblock {\em Natl. Sci. Rev.}, 5(2):203--215, 2017.

\bibitem{D30}
Paul~AM Dirac.
\newblock Note on exchange phenomena in the thomas atom.
\newblock In {\em Mathematical Proceedings of the Cambridge Philosophical
  Society}, volume~26, pages 376--385. Cambridge University Press, 1930.

\bibitem{VWN80}
Seymour~H Vosko, Leslie Wilk, and Marwan Nusair.
\newblock Accurate spin-dependent electron liquid correlation energies for
  local spin density calculations: a critical analysis.
\newblock {\em Can. J. Phys.}, 58(8):1200--1211, 1980.

\bibitem{PBE96}
John~P Perdew, Kieron Burke, and Matthias Ernzerhof.
\newblock Generalized gradient approximation made simple.
\newblock {\em Phys. Rev. Lett.}, 77(18):3865, 1996.

\bibitem{SDC94}
P.~J. Stephens, F.~J. Devlin, C.~F. Chabalowski, and M.~J. Frisch.
\newblock Ab initio calculation of vibrational absorption and circular
  dichroism spectra using density functional force fields.
\newblock {\em J. Phys. Chem.}, 98(45):11623--11627, 1994.

\bibitem{ZT08}
Yan Zhao and Donald~G Truhlar.
\newblock The m06 suite of density functionals for main group thermochemistry,
  thermochemical kinetics, noncovalent interactions, excited states, and
  transition elements: two new functionals and systematic testing of four
  m06-class functionals and 12 other functionals.
\newblock {\em Theor. Chem. Acc.}, 120(1-3):215--241, 2008.

\bibitem{PT11}
Roberto Peverati and Donald~G Truhlar.
\newblock Improving the accuracy of hybrid meta-gga density functionals by
  range separation.
\newblock {\em J. Phys. Chem. Lett.}, 2(21):2810--2817, 2011.

\bibitem{HHLT16}
S~Yu Haoyu, Xiao He, Shaohong~L Li, and Donald~G Truhlar.
\newblock Mn15: A kohn--sham global-hybrid exchange--correlation density
  functional with broad accuracy for multi-reference and single-reference
  systems and noncovalent interactions.
\newblock {\em Chem. Sci.}, 7(8):5032--5051, 2016.

\bibitem{G16}
M.~J. Frisch, G.~W. Trucks, H.~B. Schlegel, G.~E. Scuseria, M.~A. Robb, J.~R.
  Cheeseman, G.~Scalmani, V.~Barone, G.~A. Petersson, H.~Nakatsuji, X.~Li,
  M.~Caricato, A.~V. Marenich, J.~Bloino, B.~G. Janesko, R.~Gomperts,
  B.~Mennucci, H.~P. Hratchian, J.~V. Ortiz, A.~F. Izmaylov, J.~L. Sonnenberg,
  D.~Williams-Young, F.~Ding, F.~Lipparini, F.~Egidi, J.~Goings, B.~Peng,
  A.~Petrone, T.~Henderson, D.~Ranasinghe, V.~G. Zakrzewski, J.~Gao, N.~Rega,
  G.~Zheng, W.~Liang, M.~Hada, M.~Ehara, K.~Toyota, R.~Fukuda, J.~Hasegawa,
  M.~Ishida, T.~Nakajima, Y.~Honda, O.~Kitao, H.~Nakai, T.~Vreven,
  K.~Throssell, J.~A. Montgomery, {Jr.}, J.~E. Peralta, F.~Ogliaro, M.~J.
  Bearpark, J.~J. Heyd, E.~N. Brothers, K.~N. Kudin, V.~N. Staroverov, T.~A.
  Keith, R.~Kobayashi, J.~Normand, K.~Raghavachari, A.~P. Rendell, J.~C.
  Burant, S.~S. Iyengar, J.~Tomasi, M.~Cossi, J.~M. Millam, M.~Klene, C.~Adamo,
  R.~Cammi, J.~W. Ochterski, R.~L. Martin, K.~Morokuma, O.~Farkas, J.~B.
  Foresman, and D.~J. Fox.
\newblock Gaussian16 {R}evision {B}.01, 2016.
\newblock Gaussian Inc. Wallingford CT.

\bibitem{D89}
Thom~H Dunning~Jr.
\newblock Gaussian basis sets for use in correlated molecular calculations. i.
  the atoms boron through neon and hydrogen.
\newblock {\em J. Chem. Phys.}, 90(2):1007--1023, 1989.

\bibitem{WD94}
David~E Woon and Thom~H Dunning~Jr.
\newblock Gaussian basis sets for use in correlated molecular calculations. iv.
  calculation of static electrical response properties.
\newblock {\em J. Chem. Phys.}, 100(4):2975--2988, 1994.

\bibitem{KDH92}
Rick~A Kendall, Thom~H Dunning~Jr, and Robert~J Harrison.
\newblock Electron affinities of the first-row atoms revisited. systematic
  basis sets and wave functions.
\newblock {\em J. Chem. Phys.}, 96(9):6796--6806, 1992.

\bibitem{WD93}
David~E Woon and Thom~H Dunning~Jr.
\newblock Gaussian basis sets for use in correlated molecular calculations.
  iii. the atoms aluminum through argon.
\newblock {\em J. Chem. Phys.}, 98(2):1358--1371, 1993.

\bibitem{WFA03}
Florian Weigend, Filipp Furche, and Reinhart Ahlrichs.
\newblock Gaussian basis sets of quadruple zeta valence quality for atoms
  h--kr.
\newblock {\em J. Chem. Phys.}, 119(24):12753--12762, 2003.

\bibitem{RTSN02}
Jonathan~C Rienstra-Kiracofe, Gregory~S Tschumper, Henry~F Schaefer, Sreela
  Nandi, and G~Barney Ellison.
\newblock Atomic and molecular electron affinities: photoelectron experiments
  and theoretical computations.
\newblock {\em Chem. Rev.}, 102(1):231--282, 2002.

\bibitem{ZLT05}
Yan Zhao, Benjamin~J Lynch, and Donald~G Truhlar.
\newblock Multi-coefficient extrapolated density functional theory for
  thermochemistry and thermochemical kinetics.
\newblock {\em Phys. Chem. Chem. Phys.}, 7(1):43--52, 2005.

\bibitem{ZGT05}
Yan Zhao, N{\'u}ria Gonz{\'a}lez-Garc{\'\i}a, and Donald~G Truhlar.
\newblock Benchmark database of barrier heights for heavy atom transfer,
  nucleophilic substitution, association, and unimolecular reactions and its
  use to test theoretical methods.
\newblock {\em J. Phys. Chem. A}, 109(9):2012--2018, 2005.

\bibitem{T15}
{TURBOMOLE V7.0 2015}, a development of {University of Karlsruhe} and
  {Forschungszentrum Karlsruhe GmbH}, 1989-2007, {TURBOMOLE GmbH}, since 2007;
  available from {\tt http://www.turbomole.com}.

\bibitem{pyscf}
Qiming Sun, Timothy~C Berkelbach, Nick~S Blunt, George~H Booth, Sheng Guo,
  Zhendong Li, Junzi Liu, James~D McClain, Elvira~R Sayfutyarova, Sandeep
  Sharma, et~al.
\newblock Pyscf: the python-based simulations of chemistry framework.
\newblock {\em Wiley Interdisciplinary Reviews: Computational Molecular
  Science}, 8(1):e1340, 2018.

\end{thebibliography}

\label{page:end}
\end{document}